\documentclass[aps,prl,twocolumn,superscriptaddress,longbibliography]{revtex4}

\usepackage{amsmath,amssymb,bm}
\usepackage{graphicx}
\usepackage{booktabs}
\usepackage{mathtools}
\usepackage[monochrome]{xcolor}
\usepackage[colorlinks=true,allcolors=black]{hyperref}

\begin{document}

\title{Wave-Energy Partition Governs Weak Collisional Damping in Cold Plasmas}

\author{Yanzeng Zhang}
\email{yzengzhang@ustc.edu.cn}
\affiliation{School of Nuclear Science and Technology, University of Science and Technology of China, Hefei, Anhui 230027, China}
\author{Xingming Zeng}
\affiliation{School of Nuclear Science and Technology, University of Science and Technology of China, Hefei, Anhui 230027, China}
\author{Xian-Zhu Tang}
\affiliation{Theoretical Division, Los Alamos National Laboratory, Los Alamos, NM 87545, USA}
\author{Ge Zhuang}
\affiliation{School of Nuclear Science and Technology, University of Science and Technology of China, Hefei, Anhui 230027, China}

%\date{\today}

\begin{abstract}
Weak dissipation can control wave propagation, mode competition, and
instability thresholds in plasmas, yet the physical origin of large
branch-to-branch differences in collisional damping is often obscured
by dielectric-tensor calculations. We show that weak collisional
damping in cold plasmas is governed by wave-energy partition. In the
one-rate cold-plasma model, the damping rate of a \textcolor{blue}{positive-energy}
collisionless
eigenmode \textcolor{blue}{equals, to leading order in weak collisions,} the collision frequency multiplied by the
fraction of the total wave energy stored in plasma motion. This result
recasts the standard perturbative damping formula into a compact and
physically transparent law, immediately explaining why field-dominated
branches such as whistlers can be much less damped than the collision
frequency, whereas quasi-electrostatic modes can exhibit damping of
comparable magnitude. Analytic examples for Langmuir, transverse
electromagnetic, whistler, and extraordinary waves show that the
energy-partition form classifies weak collisional damping across
distinct branches and provides a simple diagnostic for mode
competition in multibranch plasma-wave systems.
\end{abstract}

\maketitle

\paragraph{Introduction.}
Weak dissipation can decisively influence wave propagation, mode
competition, and instability thresholds in plasmas. In laboratory,
space, and astrophysical environments, collisions are often weak
enough that the mode structure is still well described by
collisionless eigenmodes, yet strong enough to determine which
branches propagate efficiently, are strongly attenuated, or can
overcome a competing instability threshold
\cite{LandauLifshitz,Stix1992,Swanson2003,Brambilla1995}. A striking feature
of this regime is that waves subject to the same nominal collision
frequency can exhibit dramatically different damping rates. Some
branches are only weakly affected by collisions, whereas others are
strongly suppressed. The conventional perturbative expression for weak
collisional damping, formulated through the anti-Hermitian part of the
dielectric tensor, is well established
\cite{Brambilla1995,BreizmanKiramov2023a,BreizmanKiramov2023b},
but it does not directly expose the physical principle behind these
large branch-to-branch differences.

Here we show that this principle is wave-energy partition. \textcolor{blue}{For
positive-energy waves} in the one-rate cold-plasma model, \textcolor{blue}{in which all participating species
are assigned the same collision rate, $\nu_s=\nu $,} the weak collisional damping rate
$\Gamma_\nu$ of a collisionless eigenmode \textcolor{blue}{equals, to leading order in weak collisions,} the collision
frequency $\nu $ multiplied by the fraction of the total wave energy ($W$)
stored in plasma motion ($E_k$),
\begin{equation}
\frac{\Gamma_\nu}{\nu }=\frac{E_k}{W}.
\label{eq:main}
\end{equation}
Equivalently, weak collisions remove only the quiver kinetic-energy
reservoir of the wave~\textcolor{blue}{\cite{ShiQinFisch2021}}, while energy stored in the electromagnetic
fields is not directly damped. This compact law immediately explains
why field-dominated branches, such as whistlers, can satisfy
$\Gamma_\nu\ll \nu $, whereas quasi-electrostatic branches can
exhibit $\Gamma_\nu\sim \nu $. More generally, it provides a
branch-classification rule for weak collisional damping and a
practical diagnostic for multibranch plasma-wave problems in which
dissipation competes with ideal drive, including recent
runaway-electron-driven applications involving whistler and slow-X
branches~\cite{FulopPokolHelanderLisak2006,AleynikovBreizman2015,BreizmanKiramov2023a,BreizmanKiramov2023b,ZhangZhangTang2025}.

The result in Eq.~(\ref{eq:main}) comes from the standard weak-collision theory but
makes the physical interpretation explicit. That reformulation is useful because it
converts a formally correct but relatively opaque damping formula into
a one-line diagnostic that is easily compared across branches with
very different field and particle-energy content. The underlying
weak-collision dielectric treatment is established
\cite{Brambilla1995}, and special-case damping results for plasma
oscillations have long been known \cite{Buti1970}. The primary advance
here is the explicit cross-branch energy-partition law and its use as
a practical diagnostic.

\textcolor{blue}{The sufficient perturbative ordering adopted throughout requires
$\Gamma_\nu\ll|\omega_r|$ and $\nu_s\ll|\omega_r|$ for every participating
species $s$, where $\omega_r$
is the real collisionless eigenfrequency. These conditions apply equally to
the standard perturbative damping formula and to its energy-partition
reformulation developed below. The selected collisionless root must
also be simple and isolated, and the response used for the
perturbative expansion must be regular. }

\paragraph{Weakly collisional cold-plasma waves.}
We consider linear cold-plasma waves in the weak-collision limit. In
the absence of collisions and other dissipation, the wave equation may
be written as
\begin{equation}
\left[k_\alpha k_\beta c^2-\delta_{\alpha\beta}k^2c^2+\omega^2\epsilon_{\alpha\beta}(\omega)\right]E_\beta=0,
\label{eq:waveeq}
\end{equation}
where $\bm{k}$ is the wavevector and $\bm{E}$ is the complex
electric-field eigenvector. \textcolor{blue}{The corresponding complex magnetic-field
perturbation is $\bm{B}_1=(c/\omega)\bm{k}\times\bm{E}$.} The cold-plasma dielectric tensor is
\cite{Stix1992}
\begin{equation}
\epsilon_{\alpha\beta}=S\,\delta_{\alpha\beta}+ig\,e_{\alpha\beta\gamma}b_\gamma+(\eta-S)b_\alpha b_\beta,
\label{eq:dielectric}
\end{equation}
\textcolor{blue}{with $\bm{b}=\bm{B}_0/B_0$, where $\bm{B}_0$ is the equilibrium
magnetic field and $B_0=|\bm{B}_0|$.} In the collisionless limit,
\textcolor{blue}{
\begin{align}
S &=1-\sum_{s=e,i}\frac{\omega_{ps}^2}{\omega^2-\omega_{cs}^2},
\\
 g&=-\sum_{s=e,i}\frac{\omega_{cs}}{\omega}\frac{\omega_{ps}^2}{\omega^2-\omega_{cs}^2},
\\
\eta &=1-\sum_{s=e,i}\frac{\omega_{ps}^2}{\omega^2}.
\end{align}}
Here $\omega_{cs}$ includes the charge sign.

\textcolor{blue}{For an independent-relaxation collisional model,
a substitution of $\omega\to\omega+i\nu_s$, where all perturbations are assumed to vary as
$\exp[i(\bm{k}\cdot\bm{x}-\omega t)]$, is applied only to the dynamical
plasma response~\cite{LandauLifshitz,Stix1992,Swanson2003}, or equivalently to its contribution $\tensor{\sigma}_s$ with
the total conductivity $\tensor{\sigma}=\sum_s\tensor{\sigma}_s$. Here, $\nu_s$ denotes an effective momentum-relaxation rate
against a prescribed stationary background. For example, for an electron-dominated cold-plasma
branch on which the ion mobility is negligible, the bulk momentum-relaxing
Coulomb contribution can be represented by electron--ion friction
against effectively stationary ions, with an effective rate
$\nu_{ei}^{\mathrm{eff}}$ obtained
from $\nu_{ei}$~\cite{Brambilla1995}.
For general cross-species Coulomb dissipation, however, the response depends on
the difference between species mobilities and cannot in general be represented by
independent frequency substitutions~\cite{Brambilla1995,BrantovBychenkovRozmus2012}. As an illustration,
Appendix~A discusses the simplest momentum-conserving cold electron-ion model, in which both species are mobile and their
responses are coupled. It turns out that the internal electron-ion friction relaxes only the
relative motion while leaving the center-of-mass momentum unchanged. Hereafter, we consider only the independent-relaxation collisional model.}

The total conductivity enters the dielectric tensor through
\begin{equation}
\tensor{\epsilon}=\tensor{I}+\frac{i4\pi}{\omega}\tensor{\sigma},
\end{equation}
\textcolor{blue}{where the factor $1/\omega$ in this relation originates
from Maxwell's equations and is not shifted.}  Writing the Hermitian dielectric as
\begin{equation}
\tensor{\epsilon}_H=\tensor{I}+\sum_s \tensor{\chi}_{H,s},
\end{equation}
the leading
anti-Hermitian contribution induced by weak collisions reads
\begin{equation}
\tensor{\epsilon}_A=\frac{i}{\omega}\sum_s \nu_s\,\frac{\partial}{\partial\omega}\left(\omega\tensor{\chi}_{H,s}\right).
\label{eq:epsA_general}
\end{equation}
\textcolor{blue}{When all participating species have the same collision rate,
$\nu_s=\nu$, one obtains}
\begin{equation}
\tensor{\epsilon}_A=\frac{i\nu}{\omega}\left[\frac{\partial}{\partial\omega}\left(\omega\tensor{\epsilon}_H\right)-\tensor{I}\right].
\label{eq:epsA_one_rate}
\end{equation}

\textcolor{blue}{Here, we denote the collision-induced frequency
correction by $\delta\omega_\nu=-i\Gamma_\nu$, where
$\Gamma_\nu>0$ is the amplitude-damping rate.}
First-order perturbation theory then gives the familiar damping formula
\textcolor{blue}{in terms of the anti-Hermitian dielectric response
\cite{LandauLifshitz,Stix1992,Swanson2003,NishikawaWakatani2000}}
\begin{equation}
\Gamma_\nu=
-i\omega
\frac{\bm{E}^*\cdot\tensor{\epsilon}_A\cdot\bm{E}}
{\bm{E}^*\cdot \frac{1}{\omega}\frac{\partial}{\partial\omega}\left(\omega^2\tensor{\epsilon}_H\right)\cdot\bm{E}}.
\label{eq:standardGamma}
\end{equation}
\textcolor{blue}{This formulation has been applied to weak collisional damping in cold plasmas
\cite{Brambilla1995,BreizmanKiramov2023a,BreizmanKiramov2023b}
and to species-resolved collisionless kinetic absorption
\cite{Quataert1998}}. It is standard, while the new insight is that
it admits a much more transparent physics interpretation.

\paragraph{Energy-partition form.}
The total wave-energy density of a collisionless mode is \cite{LandauLifshitz,Stix1992}
\begin{equation}
W=\frac{1}{16\pi\omega}\bm{E}^*\cdot\frac{\partial}{\partial\omega}\left(\omega^2\tensor{\epsilon}_H\right)\cdot\bm{E}.
\label{eq:Wtotal}
\end{equation}
This energy can be decomposed into field and plasma kinetic contributions,
\begin{equation}
W= E_f+E_k=E_f+\sum_s E_{k,s}.
\label{eq:Wdecomp}
\end{equation}
The field part is
\begin{equation}
  E_f=\frac{1}{16\pi}\left(\bm{E}^*\cdot\bm{E}+\textcolor{blue}{\bm{B}_1^*\cdot\bm{B}_1}\right)
  = \frac{1}{16\pi} \bm{E}^*\cdot \left(\tensor{I} + \tensor{\epsilon}_H\right)\cdot \bm{E},
\label{eq:Ef}
\end{equation}
and the kinetic contribution of species $s$ is
\begin{equation}
E_{k,s}=\frac{1}{16\pi}\bm{E}^*\cdot\frac{\partial}{\partial\omega}\left(\omega\tensor{\chi}_{H,s}\right)\cdot\bm{E}.
\label{eq:Eks}
\end{equation}
\textcolor{blue}{Summing Eq.~\eqref{eq:Eks} over the participating species gives}
\begin{equation}
E_k=\frac{1}{16\pi}\bm{E}^*\cdot
\left[\frac{\partial}{\partial\omega}\left(\omega\tensor{\epsilon}_H\right)-\tensor{I}\right]\cdot\bm{E}.
\label{eq:Ek}
\end{equation}

\textcolor{blue}{Within the independent-relaxation model, substituting
Eq.~\eqref{eq:epsA_general} into Eq.~\eqref{eq:standardGamma} and comparing
with Eqs.~\eqref{eq:Wtotal} and \eqref{eq:Eks} yields the species-resolved
relation.}
\begin{equation}
\frac{dW}{dt}=-2\Gamma_\nu W=-\sum_s 2\nu_s E_{k,s}.
\label{eq:general_theorem}
\end{equation}
\textcolor{blue}{Here $2\Gamma_\nu$ is the wave-energy decay rate.}
\textcolor{blue}{For the one-rate case $\nu_s=\nu$, Eq.~\eqref{eq:general_theorem} reduces to}
\begin{equation}
-2\Gamma_\nu W=-2\nu  E_k.
\end{equation}
\textcolor{blue}{Consequently,}
\begin{equation}
\frac{\Gamma_\nu}{\nu }
=
\frac{E_k}{W}
=
\omega\,
\frac{
\bm{E}^* \cdot
\left[
\displaystyle \frac{\partial(\omega \tensor{\epsilon}_{H})}{\partial\omega}
-
\tensor{I}
\right]
\cdot \bm{E}
}{
\bm{E}^* \cdot
\displaystyle \frac{\partial(\omega^2 \tensor{\epsilon}_{H})}{\partial\omega}
\cdot \bm{E}
}.
\label{eq:ratio}
\end{equation}
\textcolor{blue}{An electron-dominated branch in which the ion response is
neglected provides a distinct asymptotic realization of the same compact
relation after identifying $\nu$ with the electron-specific effective
momentum-relaxation rate $\nu_{ei}^{\mathrm{eff}}$ and taking
$E_k\simeq E_{k,e}$.}
Equation~\eqref{eq:ratio} is algebraically equivalent to the standard
perturbative result, but physically much more revealing: the damping
rate is the collision frequency multiplied by the fraction of the wave
energy contained in plasma motion.

\textcolor{blue}{The implications are immediate. For positive-energy modes
with $W>0$, positive $E_{k,s}$ and $\nu_s$ in
Eq.~\eqref{eq:general_theorem} imply $\Gamma_\nu>0$ and hence collisional
damping.}
More importantly, it shows why the damping rate can be much smaller than the
underlying collision frequency. Field-dominated
waves satisfy $E_k\ll W$ and therefore $\Gamma_\nu\ll \nu $ in the one-rate limit, even
when collisions are not negligible on the particle
timescale. Conversely, quasi-electrostatic branches can have
$E_k/W=O(1)$ and thus damp at a substantial fraction of the collision
frequency, $\Gamma_\nu \sim \nu .$ 

\paragraph{Representative branches.}
The usefulness of Eq.~\eqref{eq:ratio} is best seen by applying it
across distinct cold-plasma branches.

\emph{Langmuir wave.} \textcolor{blue}{For the electrostatic Langmuir mode
propagating parallel to the equilibrium magnetic field, with
$\bm{E}\parallel\bm{k}\parallel\bm{B}_0$ and ion motion neglected, the relevant
longitudinal dielectric eigenvalue is $\eta$, and the dispersion relation reads
$\eta=1-\omega_{pe}^2/\omega^2=0$.} The standard energy
equipartition between oscillatory particle motion and electric field
gives $W=2E_k$, so that
\begin{equation}
\frac{\Gamma_\nu}{\nu }=\frac{1}{2}.
\end{equation}
This is the canonical example of a branch whose collisional damping is
an order-unity fraction of the collision frequency. It is also
consistent with classic plasma-oscillation damping results
\cite{Buti1970}.

\emph{Unmagnetized transverse electromagnetic wave.} For a transverse
electron mode in an unmagnetized plasma, \textcolor{blue}{the dielectric tensor $\tensor{\epsilon}$ can be described by the scalar $\eta $ with $\tensor{\epsilon} = \eta \tensor{I}$, where $S = \eta$ and $g=0$. As a result, the dispersion relation is 
\begin{equation}
\frac{k^2c^2}{\omega^2}=\eta=1-\frac{\omega_{pe}^2}{\omega^2},
\end{equation}}
and Eq.~\eqref{eq:ratio} leads to
\begin{equation}
\frac{\Gamma_\nu}{\nu }=\frac{\omega_{pe}^2}{2\omega^2}.
\end{equation}
At high frequency the wave becomes nearly vacuum-like, and only a
small fraction of the total energy resides in electron motion. Its
damping is therefore weak.

\begin{figure}[htbp]
\centering
\includegraphics[width=0.7\linewidth]{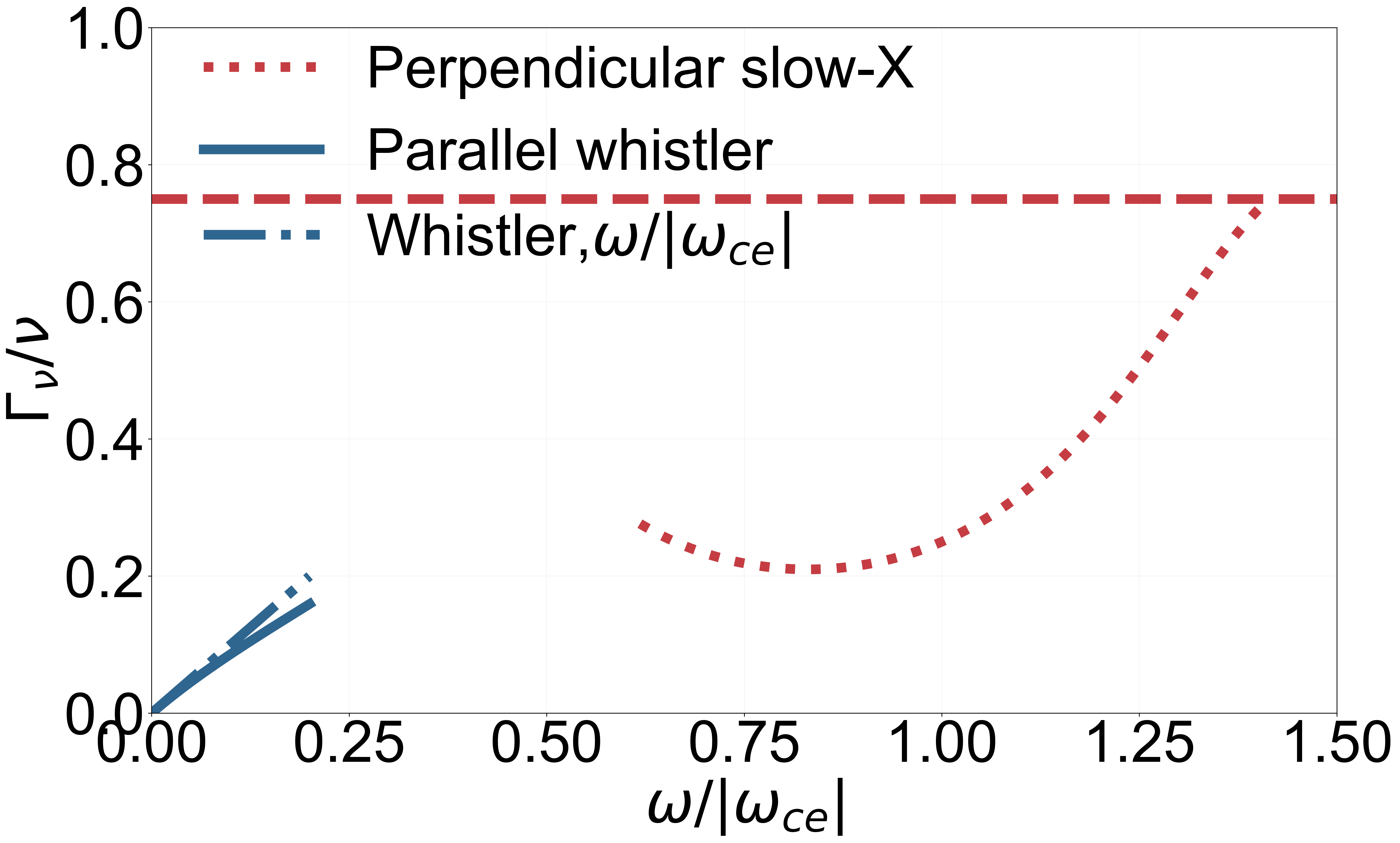}
  \includegraphics[width=0.7\linewidth]{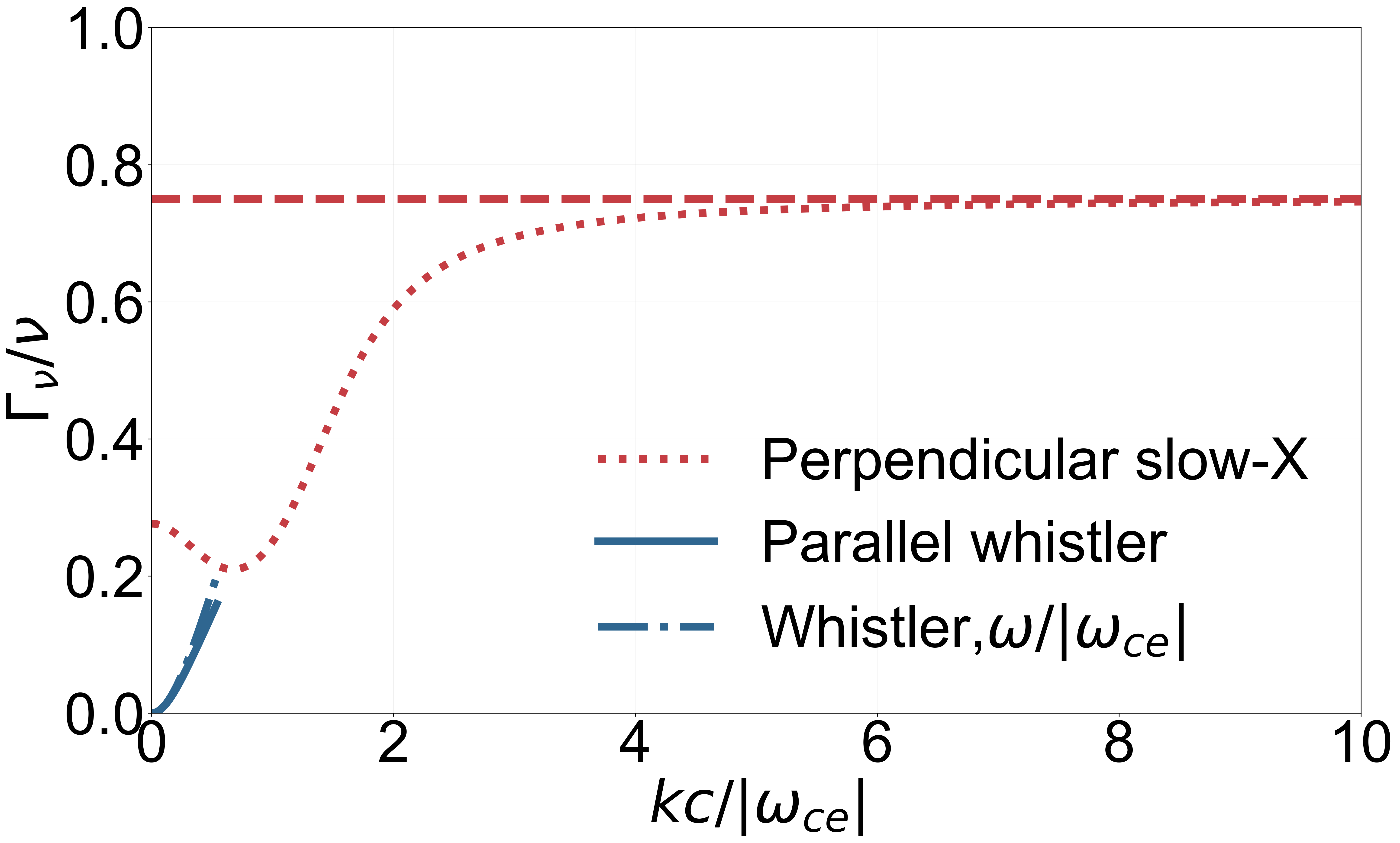}%
\caption{Numerical evaluation of Eq.~(\ref{eq:ratio}) for the parallel whistler (solid-blue) modes with $\omega<0.2|\omega_{ce}|$ and the perpendicular slow-X modes (dotted-red), where $\omega_{pe}=|\omega_{ce}|.$ The horizontal red dashed line is the slow-X damping ratio in Eq.~(\ref{eq:slowx}) and the blue dotted-dashed line is $\omega/|\omega_{ce}|$ in Eq.~\eqref{eq:whistler}. Notice that at $\omega=|\omega_{ce}|$, although $S$ and $g$
individually have poles there, these cancel in the dispersion relation
\eqref{eq:Xdisp} so that the energy and damping expressions must
be evaluated by analytic cancellation followed by the continuous
selected-mode limit rather than by a termwise expansion of the divergent dielectric
elements. }
\label{fig:whistler-slowx-validation}
\end{figure}

\emph{Parallel whistler wave.} For parallel propagation with ion dynamics neglected, in the regime \textcolor{blue}{$\omega_{ci}\ll\omega\ll |\omega_{ce}|$} and $\omega_{pe}^2\gg \omega|\omega_{ce}|$, \textcolor{blue}{the dispersion relation is characterized by the scalar dielectric eigenvalue $R\equiv S-g$ so that
\begin{equation}
\frac{k^2c^2}{\omega^2} =R 
=1-\frac{\omega_{pe}^2}{\omega(\omega-|\omega_{ce}|)}.
\end{equation}}
\textcolor{blue}{Equation (19) can then be written in terms of \(R\) as
\begin{equation}
\frac{\Gamma_\nu}{\nu }
=
\frac{
\dfrac{\partial}{\partial\omega}(\omega R)-1
}{\dfrac{1}{\omega}
\dfrac{\partial}{\partial\omega}(\omega^2 R)
}\simeq \frac{\omega}{|\omega_{ce}|}
\simeq \frac{k^2c^2}{\omega_{pe}^2}\ll 1,
\label{eq:whistler}
\end{equation}
where the numerical solutions are shown in Fig.~\ref{fig:whistler-slowx-validation}.}
This is a particularly transparent example of field-dominated energy
storage. In the whistler regime,
the wave energy density is
\begin{equation}
W = \frac{|\bm{E}|^2}{16\pi} \left[1 + R + \frac{\omega_{pe}^2}{\left(\omega - |\omega_{ce}|\right)^2}\right].
\end{equation}  
The magnetic contribution, from
$R=1-\omega_{pe}^2/\omega(\omega-|\omega_{ce}|) \approx
\omega_{pe}^2/\left(\omega|\omega_{ce}|\right),$ dominates the electric (\textcolor{blue}{the first term in the bracket}) and
kinetic (\textcolor{blue}{the third term in the bracket}) contributions, so only a small fraction of the wave energy
resides in the collisional plasma reservoir. This is precisely the
kind of branch competition emphasized in runaway-electron studies
\cite{AleynikovBreizman2015,BreizmanKiramov2023a,ZhangZhangTang2025}.

\emph{Perpendicular extraordinary mode.} \textcolor{blue}{We consider the
electron-frequency slow-X branch with ion dynamics neglected, for which
$\omega\gg\omega_{pi},\omega_{ci}$.} For perpendicular
propagation with $\bm{k}\parallel \hat{\bm{x}}$ and
\textcolor{blue}{$\bm{B}_0\parallel \hat{\bm{z}}$}, the extraordinary mode satisfies
\begin{equation}
\frac{k^2c^2}{\omega^2}=\frac{S^2-g^2}{S},
\label{eq:Xdisp}
\end{equation}
with polarization relation
\begin{equation}
\textcolor{blue}{E_y=i\frac{S}{g}E_x,}
\end{equation}
and magnetic perturbation
\begin{equation}
\textcolor{blue}{\bm{B}_1=i\frac{kc}{\omega}\frac{S}{g}E_x\hat{\bm{z}}.}
\end{equation}
\textcolor{blue}{Using the general energy formula,
one finds}
\begin{equation}
\textcolor{blue}{W=
\frac{|E_x|^2}{16\pi\omega}
\left[
\left(1+\frac{S^2}{g^2}\right)
\frac{\partial(\omega^2 S)}{\partial\omega}
-
2\frac{S}{g}\frac{\partial(\omega^2 g)}{\partial\omega}
\right],}
\end{equation}
and
\begin{equation}
\textcolor{blue}{E_k=
\frac{|E_x|^2}{16\pi}
\left[
\left(1+\frac{S^2}{g^2}\right)
\left(
\frac{\partial(\omega S)}{\partial\omega}-1
\right)
-
2\frac{S}{g}\frac{\partial(\omega g)}{\partial\omega}
\right].}
\end{equation}
Therefore,
\begin{equation}
\textcolor{blue}{\frac{\Gamma_\nu}{\nu }
=
\frac{E_k}{W}
=
\omega\,
\frac{
\left(1+\frac{S^2}{g^2}\right)
\left(
\frac{\partial(\omega S)}{\partial\omega}-1
\right)
-
2\frac{S}{g}\frac{\partial(\omega g)}{\partial\omega}
}{
\left(1+\frac{S^2}{g^2}\right)
\frac{\partial(\omega^2 S)}{\partial\omega}
-
2\frac{S}{g}\frac{\partial(\omega^2 g)}{\partial\omega}
}.}
\label{eq:Xratio}
\end{equation}

Near the slow-X resonance, where $S\rightarrow 0$ ($k\rightarrow \infty$) as
$\omega\rightarrow \omega_{UH}=\sqrt{\omega_{pe}^2 + \omega_{ce}^2},$
$|S|\ll |g|$, so the mode becomes strongly polarized along
$\bm{k}$ as $|E_x|\gg |E_y|$ and therefore more quasi-electrostatic.
\textcolor{blue}{With the $|E_x|^2$ normalization, the terms proportional to
$S/g$ and $(S/g)^2$ vanish smoothly but
$\partial S/\partial\omega\rightarrow 2\omega_{UH}/\omega_{pe}^2$ at $S\rightarrow 0$. Therefore,
both $W$ and $E_k$ remain finite.} In the limit of $|S/g|\ll 1$ and
$\omega\simeq\omega_{UH}\gg\omega_{pi},|\omega_{ci}|$, Eq.~\eqref{eq:Xratio}
yields
\begin{equation}
\frac{\Gamma_\nu}{\nu }
\approx
\omega\frac{\partial(\omega S)/\partial \omega - 1}
     {\partial(\omega^2 S)/\partial \omega},
\end{equation}
which simplifies to, with $S\approx 0$ and $\omega\approx \omega_{UH}$, 
\begin{equation}
\frac{\Gamma_\nu}{\nu }
\approx 1-
\frac{\omega_{pe}^2}{2(\omega_{pe}^2+\omega_{ce}^2)}.
\label{eq:slowx}
\end{equation}
\textcolor{blue}{At the opposite end of the same branch, $k\rightarrow0$,
the slow-X cutoff is $L=S+g=0$, with the cutoff frequency given by}
\begin{equation}
\textcolor{blue}{
\omega_L=\frac{\sqrt{\omega_{ce}^2+4\omega_{pe}^2}-|\omega_{ce}|}{2}.}
\label{eq:slowx_cutoff_frequency}
\end{equation}
\textcolor{blue}{Thus, the total wave energy and kinetic energy read}
\begin{equation}
\textcolor{blue}{
W_L=\frac{|E_x|^2}{8\pi}
\frac{2\omega_L+|\omega_{ce}|}{\omega_L+|\omega_{ce}|}>0,
~
E_{k,L}=\frac{|E_x|^2}{8\pi}
\frac{\omega_L}{\omega_L+|\omega_{ce}|},}
\end{equation}
\textcolor{blue}{and hence,}
\begin{equation}
\textcolor{blue}{
\frac{\Gamma_\nu}{\nu}
=\frac{\omega_L}{2\omega_L+|\omega_{ce}|}
=\frac{1}{2}\left(1-\frac{|\omega_{ce}|}
{\sqrt{\omega_{ce}^2+4\omega_{pe}^2}}\right).}
\label{eq:slowx_cutoff_damping}
\end{equation}
\textcolor{blue}{In the fusion-relevant dense-plasma regime
$\omega_{pe}\gtrsim|\omega_{ce}|$, Eq.~\eqref{eq:slowx_cutoff_damping} gives}
\begin{equation}
\textcolor{blue}{
0.276\lesssim
\frac{\Gamma_\nu}{\nu}
<0.5.}
\end{equation}
\textcolor{blue}{Thus, the slow-X cutoff damping rate is an $O(1)$ fraction
of $\nu$ throughout this regime. Only in the strongly magnetized
underdense limit $\omega_{pe}\ll |\omega_{ce}|$ does the ratio become
parametrically small. These two limits of $\Gamma_\nu/\nu$ at the two ends of the branch are consistent with the numerical solutions of Eq.~(\ref{eq:Xratio}) as shown in Fig.~\ref{fig:whistler-slowx-validation}.}
To sum, the slow-X branch can be damped at a substantial fraction of the collision frequency, in strong contrast with the whistler branch.

\iffalse
For convenience, these representative results are summarized in Table~\ref{tab:examples}.
\begin{widetext}
\begin{table}[b]
\caption{Representative weak-collisional damping ratios in the energy-partition form.}
\label{tab:examples}
\begin{ruledtabular}
\begin{tabular}{lll}
Wave & Collisionless dispersion & $\Gamma_\nu/\nu $ \\
\midrule
Langmuir & $1-\omega_{pe}^2/\omega^2=0$ & $1/2$ \\
Unmagnetized EM & $\frac{k^2c^2}{\omega^2}=1-\frac{\omega_{pe}^2}{\omega^2}$ & $\frac{\omega_{pe}^2}{2\omega^2}$ \\
Parallel whistler & $\frac{k^2c^2}{\omega^2}=1-\frac{\omega_{pe}^2}{\omega(\omega-|\omega_{ce}|)}$ & $\frac{\omega}{|\omega_{ce}|}$ \\
Perpendicular slow-X & $\frac{k^2c^2}{\omega^2}=\frac{S^2-g^2}{S}$ & finite, often $O(1)$ \\
\end{tabular}
\end{ruledtabular}
\end{table}
\end{widetext}

\fi

\begin{table}[b]
\caption{Representative weak-collision damping ratios in the energy-partition form. }
\label{tab:examples}
\begingroup
\renewcommand{\arraystretch}{2.5}
\begin{ruledtabular}
\begin{tabular}{lll}
Wave & Collisionless dispersion & $\Gamma_\nu/\nu $ \\
\midrule
Langmuir
& $1-\dfrac{\omega_{pe}^2}{\omega^2}=0$
& $\dfrac{1}{2}$ \\
Unmag. EM
& $\dfrac{k^2c^2}{\omega^2}
   =1-\dfrac{\omega_{pe}^2}{\omega^2}$
& $\dfrac{\omega_{pe}^2}{2\omega^2}$ \\
Para. whistler
& $\dfrac{k^2c^2}{\omega^2}
   =1-\dfrac{\omega_{pe}^2}
   {\omega(\omega-|\omega_{ce}|)}$
& $\simeq \dfrac{\omega}{|\omega_{ce}|}$ \\
Perp. slow-X
& $\dfrac{k^2c^2}{\omega^2}
   =\dfrac{S^2-g^2}{S}$
& finite, often $O(1)$ \\
\end{tabular}
\end{ruledtabular}
\endgroup
\end{table}

\paragraph{Discussion.}
Equation~\eqref{eq:ratio} provides a compact diagnostic for branch
selection in multibranch plasma-wave systems. Once a collisionless
eigenvector is known, one can evaluate $E_k/W$ directly, without
solving the full collisional dispersion relation, and thereby estimate
which branches are weakly or strongly damped. This is especially
attractive in problems where several ideal or weakly driven branches
compete and small differences in collisional damping control the
observable outcome. \textcolor{blue}{One important application concerns
runaway-electron-driven waves, where for both whistler and slow-X branches, the net linear growth rate can be estimated as
\begin{equation}
\gamma_{{\rm net}}(k,\theta)
\simeq \gamma_{{\rm RE}}(k,\theta)
-\nu(n_e,T_e)\frac{E_{k}(k,\theta)}{W(k,\theta)}.
\label{eq:runaway_net}
\end{equation}
Here $\gamma_{\rm RE}$ is obtained from the separate collisionless kinetic
susceptibility of the runaway population, whereas the collisional term acts
only on the cold bulk-plasma response. The two first-order corrections are
combined about the same collisionless cold-plasma root, whose frequency and
polarization are unchanged to leading order.
Here $\theta$ is the propagation angle with respect to the equilibrium magnetic field $\bm{B}_0$. As a result, for prescribed plasma parameters and
a runaway distribution, the largest collision rate predicted
to permit instability within the weak-drive, weak-collision regime is
\begin{equation}
\nu_{{\rm crit}}
=\max_{k,\theta}
\left[
\frac{\gamma_{{\rm RE}}(k,\theta)}
{E_{k}(k,\theta)/W(k,\theta)}
\right],
\label{eq:runaway_threshold}
\end{equation}
Here the maximization is restricted to positive-energy modes
with positive drive for which the collisionless root is simple and isolated
and the local perturbative expansion remains valid. Both the drive
$\gamma_{\rm RE}$ and the inferred collision rate must be much smaller than $|\omega_r|$.
A small $E_k/W$ does not justify extrapolating this first-order threshold
to collision rates outside that regime.
Conversely, when the shape of the runaway distribution is prescribed and its
drive scales with the runaway density, the same condition determines the
critical runaway-electron density, or corresponding critical runaway current
density, required to overcome collisional damping
\cite{ZhangZhangTang2025}. Optimization over both wavenumber and propagation
angle is important because the most unstable whistler and slow-X modes can be
oblique and can occur at different $(k,\theta)$. Notice that the purpose of the present
work is not to provide an exhaustive numerical map of $E_{k}/W$ over the
full $(k,\theta)$ space. Rather, it is to establish the energy-partition form
of collisional damping and explain why $\Gamma_{\nu}/\nu$ can differ
substantially among wave branches. A concrete branch-resolved calculation of
$\Gamma_\nu/\nu$ for runaway-resonant whistler and magnetized
electron-plasma waves is given in Ref.~\cite{AleynikovBreizman2015}.}

The result also clarifies the physical origin of a familiar but
sometimes under-explained fact: a collision frequency does not by
itself determine a wave damping rate. The relevant question is not
only how often particles collide, but how much of the wave energy
resides in the collisional plasma reservoir. That distinction is
decisive when comparing quasi-electrostatic branches to
magnetic-field-dominated ones.

The present formulation is intentionally limited.
It applies to linear cold-plasma waves in the weak-collision
regime. \textcolor{blue}{The compact relation in Eq.~\eqref{eq:main} relies on the
one-rate independent-relaxation model. For multiple rates within the same
model, Eq.~\eqref{eq:general_theorem} is the corresponding species-resolved
relation, but it is not a general Coulomb-collision theorem. A representative
cold two-fluid closure with scalar collision rates is treated in Appendix~A,
where electron--ion friction couples the two species responses. For a kinetic
Coulomb operator, however, the dissipated power must still be evaluated from
the full coupled collision operator or dielectric response
\cite{Brambilla1995,BrantovBychenkovRozmus2012}, and indeed collisional
damping in those regimes can exhibit additional structure not captured
by the present cold-plasma model~\cite{TigikZiebellYoon2016,DeVadderDeJongheKeppens2024}.} The paper also touches a
broader question concerning negative-energy waves. That connection is
physically suggestive, but negative-energy-wave behavior is not, in
general, exhausted by the present energy-partition argument alone
\cite{LashmoreDavies2007,ONeil2019}. Even so, within its natural
domain the result gives a unifying interpretation that appears to have
been underemphasized in previous presentations of weak collisional
damping.

\paragraph{Conclusion.}
We have shown that weak collisional damping of cold-plasma waves
admits a compact energy-partition form \textcolor{blue}{within the
independent-relaxation model}. In the one-rate limit, the
damping rate is the collision frequency multiplied by the fraction of
the total wave energy residing in plasma motion. \textcolor{blue}{Within
first-order weak-collision theory, this}
reformulation of standard perturbative damping theory exposes the
organizing physics of weak collisional dissipation and immediately
explains why field-dominated waves such as whistlers can remain much
less damped than the collision frequency, whereas quasi-electrostatic
branches can be damped at an order-unity fraction of it. Analytic
examples for Langmuir, transverse electromagnetic, whistler, and
extraordinary waves demonstrate that the energy-partition viewpoint
unifies standard branches and provides a practical diagnostic for mode
competition in multibranch plasma-wave systems.

\paragraph{Acknowledgments.}
This work was supported by the National MCF Energy R\&D Program of China (with Grant No. 2025YFE03200700). X.-Z. T. was supported by the U.S. Department of Energy under the Tokamak Disruption Simulation SciDAC project, jointly funded by the Office of Fusion Energy Sciences and the Office of Advanced Scientific Computing Research, both under contract
No. 89233218CNA000001.
\bibliography{collisional_damping}

\appendix
\setcounter{equation}{0}
\renewcommand{\theequation}{A\arabic{equation}}
\section{\textcolor{blue}{Appendix A: Momentum-conserving electron--ion friction}}

\textcolor{blue}{We now contrast the independent-relaxation model with the
simplest momentum-conserving electron--ion closure. Consider a uniform,
singly ionized cold plasma with the Fourier convention used in the
main text. The linear friction forces and momentum-conservation constraint are}
{\color{blue}
\begin{align}
\bm R_e&=-m_en_0\nu_{ei}(\bm u_e-\bm u_i),
\notag\\
\bm R_i&=-m_in_0\nu_{ie}(\bm u_i-\bm u_e),
\notag\\
m_en_0\nu_{ei}&=m_in_0\nu_{ie}\equiv\alpha,
\qquad
\nu_r\equiv\nu_{ei}+\nu_{ie}.
\label{eq:app_friction}
\end{align}
Thus, $\bm R_e=-\bm R_i$ and the collision frequency in the
ion equation is not equal to that in the electron equation. Introduce the
cross-product tensor $\tensor{\mathcal{C}}\cdot\bm a\equiv\bm a\times\bm b$. The two
momentum equations form the coupled block system}
{\color{blue}
\begin{align}
\left[(\nu_{ei}-i\omega)\tensor{I}-\omega_{ce}\tensor{\mathcal{C}}\right]
\cdot\bm u_e
&=-\frac{e}{m_e}\bm E+\nu_{ei}\bm u_i,
\notag\\
\left[(\nu_{ie}-i\omega)\tensor{I}-\omega_{ci}\tensor{\mathcal{C}}\right]
\cdot\bm u_i
&=\frac{e}{m_i}\bm E+\nu_{ie}\bm u_e.
\label{eq:app_block}
\end{align}}

\textcolor{blue}{Using $m_e\nu_{ei}=m_i\nu_{ie}$ and
$m_e\omega_{ce}=-m_i\omega_{ci}$, define}
{\color{blue}
\begin{align}
\tensor{\mathcal{K}}={}&\omega(\omega+i\nu_r)\tensor{I}
-i\omega(\omega_{ce}+\omega_{ci})\tensor{\mathcal{C}}
-\omega_{ce}\omega_{ci}\tensor{\mathcal{C}}^2.
\label{eq:app_K}
\end{align}}
\textcolor{blue}{Direct inversion of Eq.~\eqref{eq:app_block} then gives the
complete vector responses}
{\color{blue}
\begin{align}
\bm u_e
&=-\frac{e}{m_e}\tensor{\mathcal{K}}^{-1}\cdot
(i\omega\tensor{I}+\omega_{ci}\tensor{\mathcal{C}})\cdot\bm E,
\label{eq:app_ue}\\
\bm u_i
&= \frac{e}{m_i}\tensor{\mathcal{K}}^{-1}\cdot
(i\omega\tensor{I}+\omega_{ce}\tensor{\mathcal{C}})\cdot\bm E,
\label{eq:app_ui}\\
\bm w\equiv\bm u_i-\bm u_e
&=\frac{ie\omega}{m_r}\tensor{\mathcal{K}}^{-1}\cdot\bm E,
\qquad
m_r\equiv\frac{m_em_i}{m_e+m_i}.
\label{eq:app_w}
\end{align}
The current $\bm J=en_0\bm w$ is therefore controlled by one
joint response tensor rather than by a sum of independently shifted species
mobilities. With $\omega_p^2\equiv\omega_{pe}^2+\omega_{pi}^2
=4\pi n_0e^2/m_r$,}
{\color{blue}
\begin{equation}
\tensor{\sigma}_c=\frac{i\omega\omega_p^2}{4\pi}\tensor{\mathcal{K}}^{-1},
\qquad
\tensor{\epsilon}_c=\tensor{I}-\omega_p^2\tensor{\mathcal{K}}^{-1}.
\label{eq:app_dielectric}
\end{equation}}
\textcolor{blue}{The Cartesian gyrotropic elements follow directly from this
tensor equation. With $\tensor{\mathcal{P}}_\parallel=\bm b\bm b$,
$\tensor{\mathcal{P}}_\perp=\tensor{I}-\bm b\bm b$,}
{\color{blue}
\begin{gather}
a_0=\omega(\omega+i\nu_r),
\qquad \mathcal A_\perp=a_0+\omega_{ce}\omega_{ci},
\notag\\
\mathcal B=\omega(\omega_{ce}+\omega_{ci}),
\qquad \Delta_\perp=\mathcal A_\perp^2-\mathcal B^2,
\notag\\
\tensor{\mathcal{K}}^{-1}=\frac{\tensor{\mathcal{P}}_\parallel}{a_0}
+\frac{1}{\Delta_\perp}
\left(\mathcal A_\perp\tensor{\mathcal{P}}_\perp
+i\mathcal B\tensor{\mathcal{C}}\right),
\label{eq:app_K_inverse}
\end{gather}}
\textcolor{blue}{so that}
{\color{blue}
\begin{align}
\tensor{\epsilon}_c
&=S_c\tensor{\mathcal{P}}_\perp
+\eta_c\tensor{\mathcal{P}}_\parallel
+i g_c\tensor{\mathcal{C}},
\label{eq:app_gyrotropic}\\
S_c&=1-\frac{\omega_p^2\mathcal A_\perp}{\Delta_\perp},
&g_c&=-\frac{\omega_p^2\mathcal B}{\Delta_\perp},
\notag\\
\eta_c&=1-\frac{\omega_p^2}{\omega(\omega+i\nu_r)}.
\label{eq:app_elements}
\end{align}}
\textcolor{blue}{Equations~\eqref{eq:app_dielectric}-\eqref{eq:app_elements}
show explicitly why the coupled response is not generated by assigning
independent frequency shifts to the electron and ion susceptibilities.}

\textcolor{blue}{We next obtain the damping directly from the complex
dispersion relation. At fixed real $\bm k$, define the wave operator}
{\color{blue}
\begin{equation}
\begin{aligned}
\tensor{\mathcal{M}}(\omega,\nu_r;\bm k)
&=c^2\bm k\bm k-c^2k^2\tensor{I}
+\omega^2\tensor{\epsilon}_c(\omega,\nu_r),
\\
\det\tensor{\mathcal{M}}&=0.
\end{aligned}
\label{eq:app_wave_operator}
\end{equation}}
\textcolor{blue}{Let $\omega_0$ be a real, simple collisionless root with
eigenvector $\bm E_0$, and write
$\omega=\omega_0-i\Gamma_\nu+O(\nu_r^2)$. First-order expansion of
$\tensor{\mathcal{M}}\cdot\bm E=0$ gives}
{\color{blue}
\begin{equation}
\Gamma_\nu
=-i\nu_r
\frac{\bm E_0^*\cdot
(\partial_{\nu_r}\tensor{\mathcal{M}})_0\cdot\bm E_0}
{\bm E_0^*\cdot
(\partial_\omega\tensor{\mathcal{M}})_0\cdot\bm E_0}.
\label{eq:app_gamma_perturb}
\end{equation}
Here the derivatives are evaluated at
$(\omega,\nu_r)=(\omega_0,0)$ and the denominator derivative is taken at
fixed $\bm k$. Writing
$\tensor{\mathcal{K}}_0\equiv\tensor{\mathcal{K}}(\omega_0,0)$, note that at
$\nu_r=0$, $\tensor{\mathcal{K}}_0$ is Hermitian, so
$(\tensor{\mathcal{K}}_0^{-1})^\dagger
\tensor{\mathcal{K}}_0^{-1}=\tensor{\mathcal{K}}_0^{-2}$. One then obtains}
{\color{blue}
\begin{align}
(\partial_{\nu_r}\tensor{\mathcal{M}})_0
&=i\omega_0^3\omega_p^2\tensor{\mathcal{K}}_0^{-2},
\label{eq:app_M_nu}\\
\bm E_0^*\cdot(\partial_\omega\tensor{\mathcal{M}})_0\cdot\bm E_0
&=16\pi\omega_0 W,
\label{eq:app_M_energy}
\end{align}
where $W$ is the total time-averaged collisionless wave
energy defined in Eq.~\eqref{eq:Wtotal}. Therefore, the complex root of the
coupled dielectric dispersion relation gives}
{\color{blue}
\begin{equation}
\Gamma_\nu^{(\mathrm{disp})}
=\nu_r\frac{\omega_0^2\omega_p^2}{16\pi W}
\bm E_0^*\cdot\tensor{\mathcal{K}}_0^{-2}\cdot\bm E_0.
\label{eq:app_gamma_disp}
\end{equation}}

\textcolor{blue}{The energy reservoir in this result can be calculated
independently. The time-averaged kinetic energy separates into center-of-mass
and relative parts, but the equal-and-opposite friction forces act only on the
latter:}
{\color{blue}
\begin{align}
E_{k,\mathrm{rel}}
&=\frac{n_0m_r}{4}|\bm u_i-\bm u_e|^2
\notag\\
&=\frac{\omega_0^2\omega_p^2}{16\pi}
\bm E_0^*\cdot\tensor{\mathcal{K}}_0^{-2}\cdot\bm E_0.
\label{eq:app_Erel}
\end{align}}
\textcolor{blue}{Direct comparison of Eqs.~\eqref{eq:app_gamma_disp} and
\eqref{eq:app_Erel} yields}
{\color{blue}
\begin{equation}
\frac{\Gamma_\nu}{\nu_r}
=\frac{E_{k,\mathrm{rel}}}{W}.
\label{eq:app_relative_partition}
\end{equation}
The same factor and sign follow from the mechanical power:
$P_{\mathrm{fric}}=\alpha|\bm u_i-\bm u_e|^2/2
=2\nu_rE_{k,\mathrm{rel}}=2\Gamma_\nu W$. Thus, unlike
Eq.~\eqref{eq:general_theorem}, internal electron-ion friction does not
damp $E_{k,e}$ and $E_{k,i}$ as independent reservoirs; it relaxes their
relative motion while leaving the center-of-mass momentum unchanged.}

\textcolor{blue}{Finally, in the formal stationary-ion limit $m_i/m_e\to\infty$,
one has $\omega_{pi}^2,\omega_{ci},\nu_{ie}\to0$ and
$\nu_r\to\nu_{ei}$. Then $\bm w\to-\bm u_e$ and}
{\color{blue}
\begin{align}
\tensor{\epsilon}_e
&=\tensor{I}-\frac{\omega_{pe}^2}{\omega}
\left[(\omega+i\nu_{ei})\tensor{I}
-i\omega_{ce}\tensor{\mathcal{C}}\right]^{-1},
\label{eq:app_stationary_epsilon}\\
\frac{\Gamma_\nu}{\nu_{ei}}
&=\frac{E_{k,e}}{W}.
\label{eq:app_stationary_partition}
\end{align}
For finite ion mass, this stationary-ion reduction requires
$|\bm u_i|\ll|\bm u_e|$ together with a negligible ion contribution to the
selected mode's dispersion relation and wave energy. The velocity ordering
alone does not ensure negligible ion kinetic energy because of its mass weighting.
The reduction fails near ion response poles or on ion-scale branches.
The scalar two-fluid result is the fluid counterpart of the
relative-mobility structure obtained from the Landau operator by
Brambilla~\cite{Brambilla1995}. }

\end{document}